\documentclass{jfm}

\usepackage{amsmath}
\usepackage{graphicx}
\usepackage{natbib}
\usepackage{color}
\usepackage{bm}

\ifCUPmtlplainloaded \else
  \checkfont{eurm10}
  \iffontfound
    \IfFileExists{upmath.sty}
      {\typeout{^^JFound AMS Euler Roman fonts on the system,
                   using the 'upmath' package.^^J}%
       \usepackage{upmath}}
      {\typeout{^^JFound AMS Euler Roman fonts on the system, but you
                   dont seem to have the}%
       \typeout{'upmath' package installed. JFM.cls can take advantage
                 of these fonts,^^Jif you use 'upmath' package.^^J}%
      }
  \else
  \fi
\fi

\ifCUPmtlplainloaded \else
  \checkfont{msam10}
  \iffontfound
    \IfFileExists{amssymb.sty}
      {\typeout{^^JFound AMS Symbol fonts on the system, using the
                'amssymb' package.^^J}%
       \usepackage{amssymb}%

      }{}
  \fi
\fi

\ifCUPmtlplainloaded \else
  \IfFileExists{amsbsy.sty}
    {\typeout{^^JFound the 'amsbsy' package on the system, using it.^^J}%
     \usepackage{amsbsy}}
    {}
\fi

\newsavebox{\astrutbox}
\sbox{\astrutbox}{\rule[-5pt]{0pt}{20pt}}

\newcommand\tieq[1]{\hspace{0.5cm} \text{#1}\hspace{0.5cm}} %text in equation environmen
\newcommand\pstat{p_{\infty}}

\newcommand{\mean}[1]{\langle {#1}\rangle}

\title[]{Turbulent pair dispersion as a continuous-time random walk}
\author[S. Thalabard, G. Krstulovic and J. Bec]%
{Simon Thalabard,\ns Giorgio Krstulovic\break and J{\'e}r{\'e}mie
  Bec\thanks{Email address for correspondence: jeremie.bec@oca.eu}}

\affiliation{Laboratoire Lagrange UMR 7293, Universit\'e de
  Nice-Sophia Antipolis,\\ CNRS, Observatoire de la C\^ote d'Azur,
  Bd.\ de l'Observatoire, 06300 Nice, France.}

\begin{document}

\maketitle

\begin{abstract}
  The phenomenology of turbulent relative dispersion is revisited. A
  heuristic scenario is proposed, in which pairs of tracers undergo a
  succession of independent ballistic separations during time
  intervals whose lengths fluctuate. This approach suggests that the
  logarithm of the distance between tracers self-averages and performs
  a continuous-time random walk. This leads to specific predictions
  for the probability distribution of separations, that differ from
  those obtained using scale-dependent eddy-diffusivity models
  (\textit{e.g.}\/ in the framework of Richardson's approach). Such
  predictions are tested against high-resolution simulations and shed
  new lights on the explosive separation between tracers.
\end{abstract}

\section{Introduction}
\label{sec:intro}

Tracers in a turbulent flow separate in an explosive manner. Their
averaged square distance becomes independent of their initial
separation and grows as $t^3$ at large times. This explains the
ability of turbulence to considerably enhance mixing
\citep{dimotakis:2005}, but also links to fundamental issues in
turbulence, where a key question is to relate the irregularity of the
Lagrangian flow with the persistence of a finite dissipation at
infinite Reynolds numbers \citep{cardy:2008,eyink-drivas:2014}. Since
the first observation of the $t^3$ law by \cite{richardson:1926} and
its interpretation in terms of Kolmogorov's similarity hypothesis by
\cite{obukhov:1941}, precise experimental and numerical measurements
and acute modelling of pair separation have proven to be a
particularly laborious exercise, as stressed for instance in the
reviews by \cite{sawford:2001} and \cite{salazar-collins:2009}.

A difficulty in observing the explosive law stems from the huge
separation of timescales that it requires.  \cite{batchelor:1950}
indeed showed that the $t^3$ law is preceded by a ballistic regime
during which the mean-square separation is $\propto t^2$.  This
dominates relative dispersion as long as the initial velocity
difference between the tracers has not changed much, that is up to
times of the order of the eddy turnover time $\tau_{r_0}\propto
r_0^{2/3}$ associated to the inertial-range initial separation $r_0$
---\,hereafter referred to as Batchelor's timescale. The explosive
$t^3$ law takes over at times $t\gg\tau_{r_0}$.  While Batchelor's
predictions are quantitatively confirmed in particle-tracking
experiments \citep{berg-etal:2006,ouellette-etal:2006} and in direct
numerical simulations
\citep{yeung:1994,sawford-etal:2008,bitane-etal:2012}, the most
manifest observations of the $t^3$ law are limited to initial
separations $r_0$ close to the Kolmogorov dissipative scale $\eta$
\citep{ott-mann:2000,boffetta-sokolov:2002b,biferale-etal:2005,eyink:2011}. For
particles whose initial separation lies in the inertial range, the
$t^3$ growth is more elusive and emerges at best as a short transient
on times both much larger than $\tau_{r_0}$ and smaller than the
integral timescale, from which the separation between tracer particles
becomes purely diffusive.  Even though the $t^3$ explosive law can be
understood on purely dimensional grounds, there is today a lack of
sufficiently accurate data that would substantiate the $t^3$ law and
constrain the underlying mechanisms.

Long-established modelling of relative dispersion is based on the
assumption that distances between tracers undergo a scale-dependent
diffusion~\citep{richardson:1926}. As stressed by
\cite{falkovich-etal:2001}, this presupposes to observe the system on
a timescale $t$ much longer than the Lagrangian correlation time
$\tau_r^\mathcal{L}$ of the velocity difference between tracers at a
distance $r$. For separations of the order of the typical separation
$r \!\sim \!  \varepsilon^{1/2}t^{3/2}$ (with $\varepsilon$ denoting
the mean kinetic energy dissipation rate of the turbulent flow), this
presumes that $\tau_r^\mathcal{L}\!\ll
\!\varepsilon^{-1/3}r^{2/3}\!\sim\!\tau_r$ and thus leads to the
unrealistic assumption that the Lagrangian correlation time is much
shorter than the Eulerian turnover time.  Markovian models involving
the joint evolution of separations and velocity differences have been
introduced to circumvent such drawbacks
\citep{kurbanmuradov-sabelfeld:1995,thomson-wilson:2013}. They are
mainly based on the shortness of acceleration correlation times in
turbulence and usually rely on the input of Eulerian single-time
statistics. Possible pitfalls in their justification relate to
neglecting long-term memory effects due to the persistence of
turbulent flows. In an ideal infinite-Reynolds number turbulent flow,
most Markovian models admit scaling solutions of the form $r\propto
t^{3/2}$ and $v=(\mathrm{d}r/\mathrm{d}t) \propto t^{1/2}$. However,
at the same time, a small-scale regularization is usually required in
order to prevent particle pairs from collapsing together at a finite
time with a vanishing velocity difference. The scaling solutions
depicted above are usually coexisting with their ``dual'' $r\propto
(t_\star-t)^{3/2}$ with $v\propto -(t_\star-t)^{1/2}$. On the one
hand, preventing possible finite-time singularities requires modelling
the dissipative-range physics of turbulence. On the other hand, such
events might correspond to the loss of memory that could justify the
applicability of diffusive approaches at sufficiently large times. The
recent numerical studies of relative dispersion by
\cite{scatamacchia-etal:2012} and \cite{bitane-etal:2013} have clearly
stressed the importance of small scales on the overall evolution of
distances: Some pairs remain trapped at scales $r\ll r_0$ for very
long times and give an important contribution to the average
separation. Accounting for such trapping events has motivated the
introduction of non-Markovian models
\citep{shlesinger-etal:1987,faller:1996,rast-pinton:2011}.  Most of
them rely on prescribing a distribution of waiting times.

We follow here a slightly different route to account for
non-Markovianity. Numerical data from a large-scale simulation of 3D
homogenous turbulence are used, to question the possibility that
systematic deviations for the distribution of the inter-particle
separations from Richardson's self-similar solution could stem from a
multiplicative process.  We motivate this possibility using a
handwaving description of a ballistic phenomenology, observed to be
compatible with the bulk of the distribution of distances.  We then
flesh this observation and describe a piecewise-ballistic
phenomenological toy model, which yields a whole family of
self-similar distributions for the inter-particle distances, and
predict that the distributions of their \emph{logarithms} should
collapse towards a well-defined distribution.  This prediction is then
tested against numerical data.

\section{The inter-particle distance as a multiplicative process?}
\label{sec:multiplicative}

Most models for relative dispersion reproduce the $t^3$ explosive law
fo the long-time behaviour of the mean-square separation.  However,
they usually yield different probability distributions of the distance
$r$ between tracers at time $t$. Richardson's eddy-diffusivity
approach leads to the self-similar distribution
\citep[see][]{salazar-collins:2009}
\begin{equation}
  p(r,t)=\frac{426}{35}\sqrt{\frac{286}{\pi}} \frac{r^2}{(g\,
    \varepsilon)^{3/2} \, t^{9/2}} \exp{\left[-\frac{1}{2} \sqrt[3]{1287} \,
      \frac{r^{2/3}}{(g\,\varepsilon)^{1/3}\,t}\right]},
  \label{Eq:PDFRichardson}
\end{equation}
which is uniquely determined by Richardson's constant $g=\langle
r^2\rangle/(\varepsilon t^3)$. The only settings where this
distribution is exact is for delta-correlated-in-time flows
\citep{falkovich-etal:2001}.  Be it using massive numerical
simulations or sophisticated particle-tracking experiments, one
expects to measure some systematic deviations from this distribution.
Those deviations exist and have been reported but, quite remarkably,
they are rather mild: Richardson's diffusive mechanism appears to
predict correctly the bulk of the distribution and only fails to
accurately describe rare events.  More precisely, for tracers with an
initial distance $r_0$ of the order of the Kolmogorov scale,
Richardson's distribution seems to overestimate the fraction of
particles separating faster than $t^3$ \citep{scatamacchia-etal:2012}.
When $r_0$ lies within the inertial range, it underestimates the
fraction of pairs that separate faster than the average
\citep{bitane-etal:2013}. As those deviations are very fine, one
cannot preclude that they are either a consequence of intermittency or
of finite-Reynolds-number effects that would contaminate the
statistics at both very large and very small scales.  A more
serious concern exists, though: As stated earlier, the
eddy-diffusivity framework entails short correlation times for the
velocity differences between tracers and has thus debatable physical
origins.

In order to remain closer to phenomenological considerations, we
propose a completely different mechanism that relies on Batchelor's
ballistic separation of pairs. Let us imagine the following simplified
scenario.  Two tracers initially separated by a distance $r_0$ inside
the inertial range will follow a ballistic motion with a velocity
difference $\delta u_0$ for a short period of time $\tau_0$. By
Kolmogorov 1941 phenomenology the time and velocity should scale as
$\tau_0\sim r_0^{2/3}$ and $\delta u_0\sim r_0^{1/3}$. Therefore,
after the time $\tau_0$ the pair will be separated by a distance
$r_{\tau_0}=r_0+\tau_0\,\delta u_0=r_0(1+a_0)$, where $a_0$ is in
principle a scale-independent random variable that accounts for the
fluctuations of the flow. Applying the same argument to $r_{\tau_0}$
we find $r_{\tau_0+\tau_1}=r_0\,(1+a_0)\,(1+a_1)$ where $a_1$ is
independent of $a_0$, and so on. Finally, the relative distance
between the two tracers at time $t$ is given by
$r_t=r_0\,(1+a_0)\cdots(1+a_n)$ with $t=\tau_0+\cdots+\tau_n$, where
by construction the number of terms is $n\sim \ln{t}$. Although naive,
such a piecewise-ballistic phenomenology relies on the physical ideas
of Batchelor and makes use of the ballistic separation between tracers
whose validity is strongly assessed by numerics and experiments. This
scenario suggests that the $t^3$ law may appear as the consequence of
the separations increasing multiplicatively rather than
additively.  If such was indeed the case, the natural observable is
not the distance but rather its \emph{logarithm}, which is
then expected to self-average. Indeed all the $n$ random
variables $a_i$ are independent and identically distributed. By the
law of large numbers, we thus expect at large times $\ln r_t \propto
n\sim \ln t $.

\begin{figure}
  \centerline{ \includegraphics[height=0.35\textwidth]{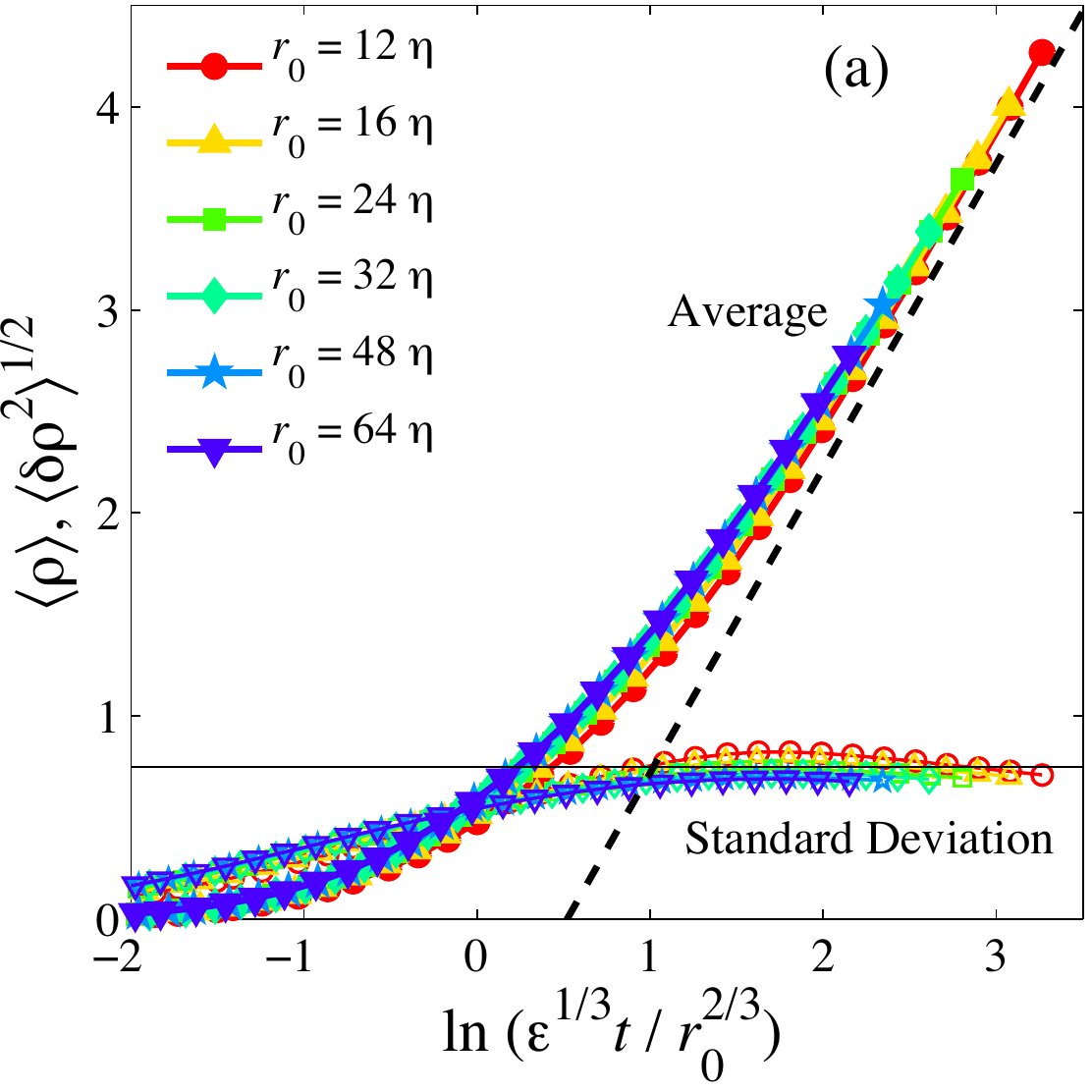}
  \qquad \includegraphics[height=0.35\textwidth]{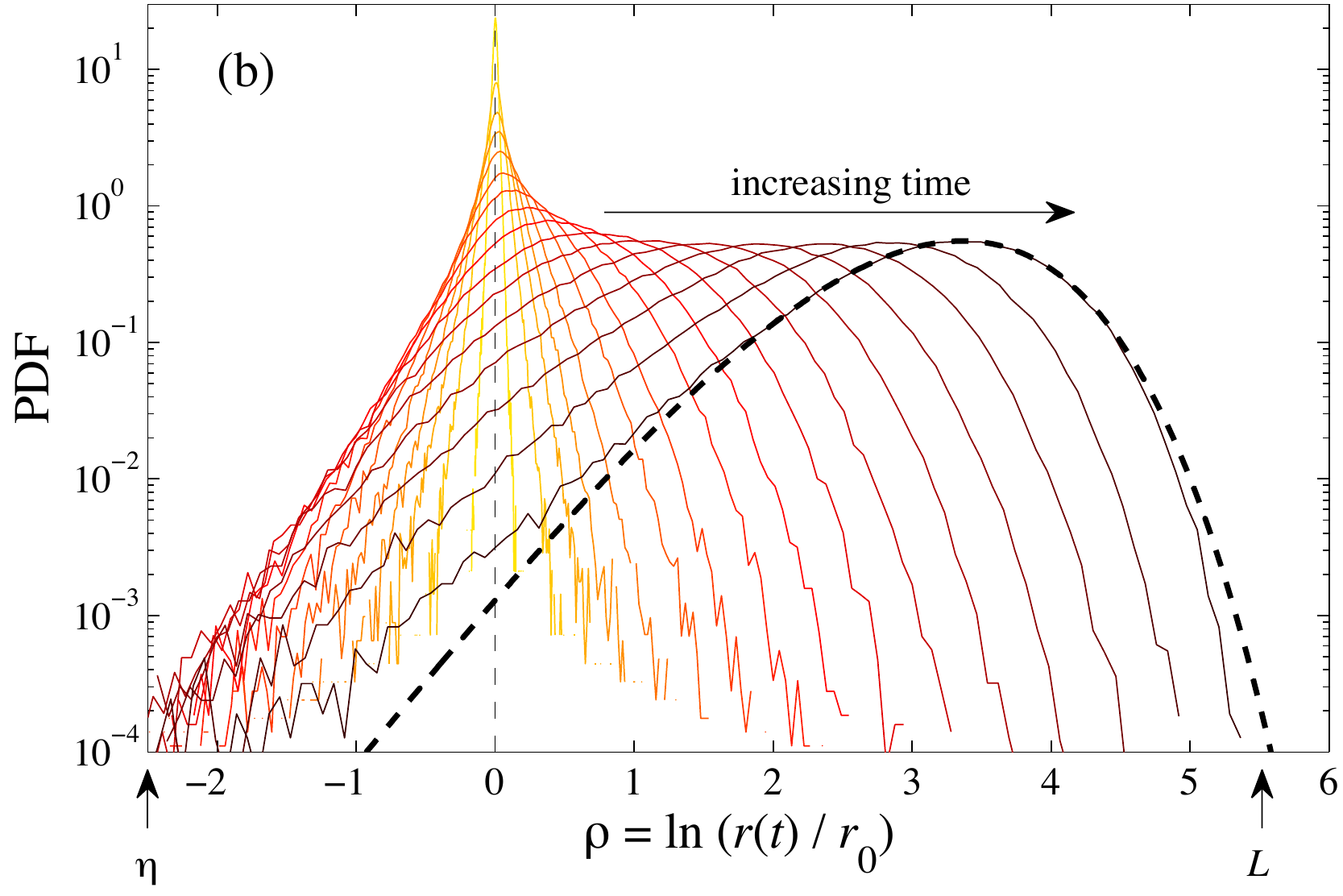} }
\caption{(Colour online) (a) Average value and standard deviation of
  $\rho = \ln (r(t)/r_0)$ for various initial separations $r_0$ and as
  a function of $\ln (t/\tau_{r_0})$ with $\tau_{r_0} =
  \varepsilon^{-1/3}r_0^{2/3}$.  The dashed line is $\langle \rho
  \rangle = ({3}/{2}) \ln(t/\tau_{r_0} )+ ({1}/{2}) \log g-0.46$ with
  $g=0.52$ and the solid line $\langle\delta\rho^2\rangle^{1/2} =
  0.748$; both correspond to Richardson's self-similar predictions.  (b)
  Distributions of $\rho$ for $r_0 = 12\eta$ and various times; the
  two vertical arrows mark the dissipative scale $\eta$ and the
  integral scale $L$. The dashed line is Richardson's distribution.}
\label{fig:evol_distrib_lnR}
\end{figure}

To test those ideas we analyse data from a large direct numerical
simulation of 3D homogeneous turbulence inside a triply-periodic box,
which uses $4096^3$ grid points to achieve a Taylor-based Reynolds
number $R_\lambda \simeq 730$ and is seeded with $10^7$ tracers
\citep[see][for more details]{bitane-etal:2013}.  Figure
\ref{fig:evol_distrib_lnR}(a) shows the evolution of the average of
the logarithm $\rho=\ln{(r_t/r_0)}$ of the inter-particle separations,
together with its standard deviation.  The statistics are conditioned
on the initial distances $r_0$ between the particles, which are chosen
to lie within the inertial range (between $12$ and $64 \eta$). To deal
with dimensionless quantities, distances between pairs of tracers are
divided by their initial separations, while timescales are divided by
the associated Batchelor time scale.
At large times the averaged logarithm of the separation exhibits a
slant asymptote, whose linear slope cannot be quantitatively
distinguished from $3/2$: The dashed line shows a behaviour
$\mean{\rho}\approx(3/2)\ln{t}+const$. From the law of large numbers,
one also expect that the variance
$\mean{\delta\rho^2}=\mean{\rho^2}-\mean{\rho}^2$ increases as
$\ln{t}$. This feature is clearly not observed: The standard deviation
of the logarithm rather reaches a plateau for times much longer than
the Batchelor time $\tau_{r_0}$.  From the latter observation, we
cannot exclude that the evolution of the inter-particle distances be
multiplicative. Yet, such a multiplicative evolution would need to be
non-Markovian. In other words a non constant time stepping may be
involved in the above phenomenology. This would break the assumptions
of the law of large numbers.

Figure \ref{fig:evol_distrib_lnR}(b) shows the time evolution of the
probability distribution of $\rho$. After a short transient, one
observes a self-similar regime, which might explain the behaviours of
both the average and the standard deviation. Indeed, if the
distribution of distances takes the scale-invariant form
$p(r,t)=r^{-1} \Phi[r/\ell(t)]$, with $\ell(t)$ an arbitrary function
of time, then the distribution of $\rho=\ln{(r/r_0)}$ can be written
$p(\rho,t)=\Psi[\rho-\log(\ell(t)/r_0)]$, with $\Psi[z]=\Phi[{\rm
  e}^z]$. This leads to
\begin{equation}
  \langle \rho\rangle=\ln(\ell(t)/r_0)+\langle z \rangle, \quad \langle
  \delta\rho^2\rangle=\langle \delta z^2\rangle\label{Eq:SelfSimPred}
\end{equation}
where the (time-independent) moments of $z$ are obtained by using the
distribution $\Psi[z]$. Note that by construction $\langle \log r
\rangle$ does not depend on the initial conditioning $r_0$. If in
addition $\ell(t)$ is a power-law of $t$, then $\langle \rho \rangle
\propto \ln t$ and $\langle \delta\rho^2 \rangle \simeq const$, as
observed in the data.  For Richardson distribution
\eqref{Eq:PDFRichardson}, $\ell(t)=(g\,\varepsilon)^{1/2}\,t^{3/2}$
and, remarkably, the full distribution of $z$ depends neither on the
constant $g$, nor on any other physical quantities. Hence, once
$\langle \rho \rangle$ is known, the distribution of $\rho$ induced
by~\eqref{Eq:PDFRichardson} does not require any fitting. The
behaviours associated to Richardson's distribution are displayed in
Figure~\ref{fig:evol_distrib_lnR}(a) ---\,dashed and solid lines. They
give a good approximation for both the mean and the standard deviation
of $\rho$. As seen in Figure~\ref{fig:evol_distrib_lnR}(b), after the
transient given by the Batchelor timescale, Richardson's distribution
reproduces well the bulk but fails to accurately predict the extreme
fluctuations of $\rho$. This can hardly be blamed on finite-range
effects as almost all separations are well inside the inertial range
---\,as indicated by the two vertical arrows.

The most noticeable departure from Richardson's distribution occurs at
separations much smaller than their average. We indeed clearly observe
$p(\rho)\propto \mathrm{e}^{2\rho}$, hereby evidencing a power-law
behaviour $p(r)\propto r$ for distances \citep[as already observed
by][]{bitane-etal:2013}. This contrasts to the scaling $p(r)\propto
r^2$ obtained from eddy-diffusivity models in the K41 framework,
including in addition to Richardson's model those where the diffusion
coefficient depends also on time \citep[\textit{i.e.}\/ of the form
$D\sim \varepsilon^\alpha r^\beta t^\gamma$,
see][]{Hentschel_Procaccia1984}. This suggests that the statistics of
extreme events cannot be captured by simple diffusive models. However,
as we will now see, the heuristic multiplicative approach described
earlier can be refined. It leads to a qualitative understanding of the
self-similarity and the tails of the distributions observed in Figure
\ref{fig:evol_distrib_lnR}(b).

\section{A piecewise-ballistic heuristic scenario}
\label{sec:ballistic}

A model that predicts the long-time $t^3$ explosive law but disregards
the short-time ballistic behaviour may yield some biased insights, as
it will probably fail to take into account the intrinsic non-Markovian
nature of pair separation.  \cite{ilyin-etal:2010} and
\cite{eyink-benveniste:2013} therefore choose to study pair dispersion
in terms of a diffusion equation with a memory kernel. We explore here
an alternative scenario where we literally implement the ballistic
ideas. We argue that this short-time behaviour can be thought of not
only as a transient feature but as the central ingredient, that yields
the explosive regime. To this end, we propose a piecewise-ballistic
scenario of tracers separation, described in terms of a toy stochastic
model which fits into the general class of so-called
\emph{Continuous-Time Random Walks} \citep[CTRW,
see][]{hughes:1995}. The model is rooted on a very intuitive
phenomenology, which is in a sense built in to yield the $t^3$ law.
Yet, it also captures non-trivial features of the large-time
statistics of the separations between tracers, among which their
self-similarity, their explosive nature, and
a qualitative description of the distributions of extreme events.

\subsection{Intuitive description of a stochastic piecewise-ballistic model}
\begin{figure}
\begin{minipage}{0.4\textwidth}
  \centerline{\includegraphics[height=0.8\textwidth]{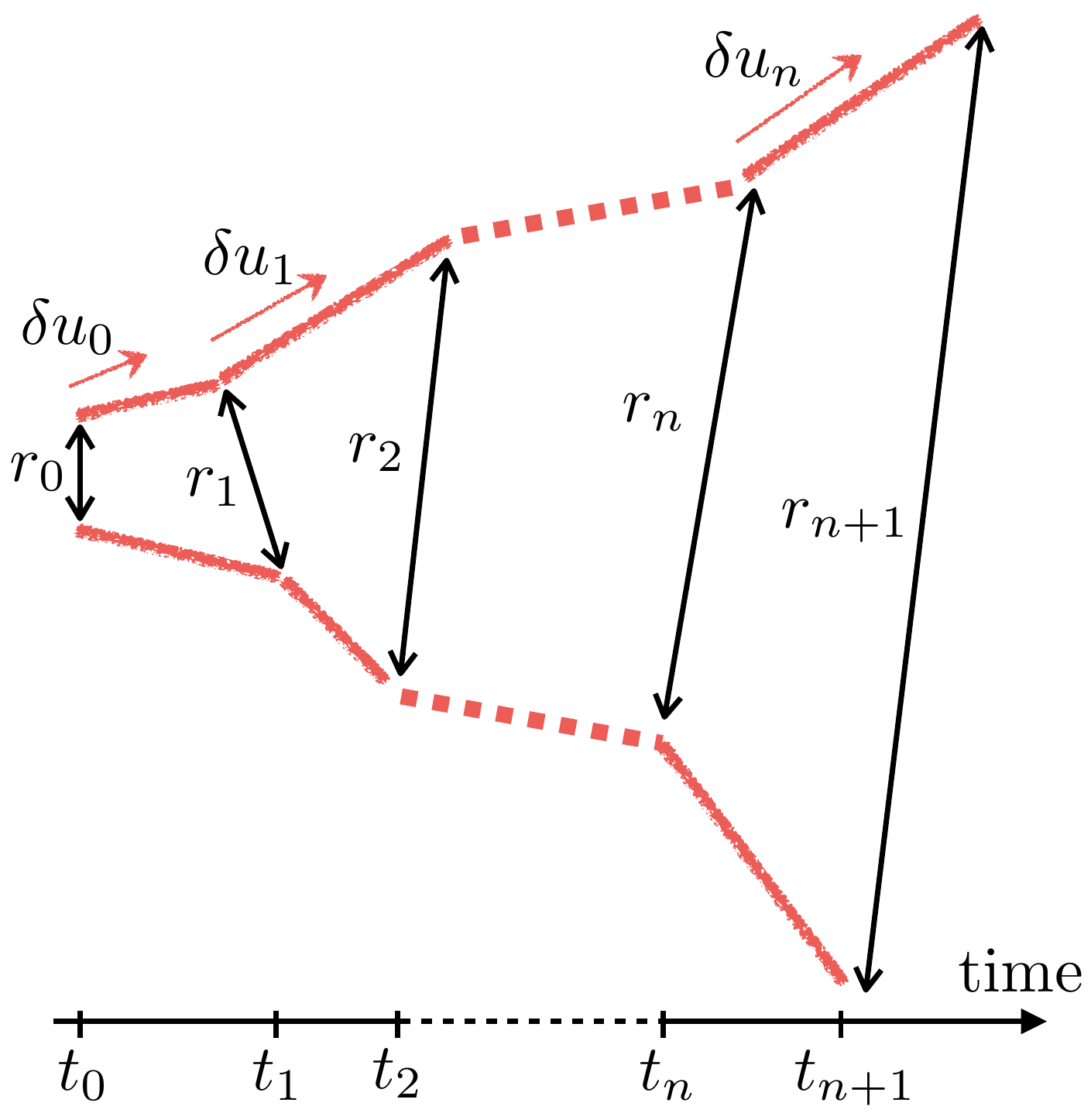}}
\end{minipage}    
\begin{minipage}{0.45\textwidth}
  \centerline{\includegraphics[height=0.8\textwidth]{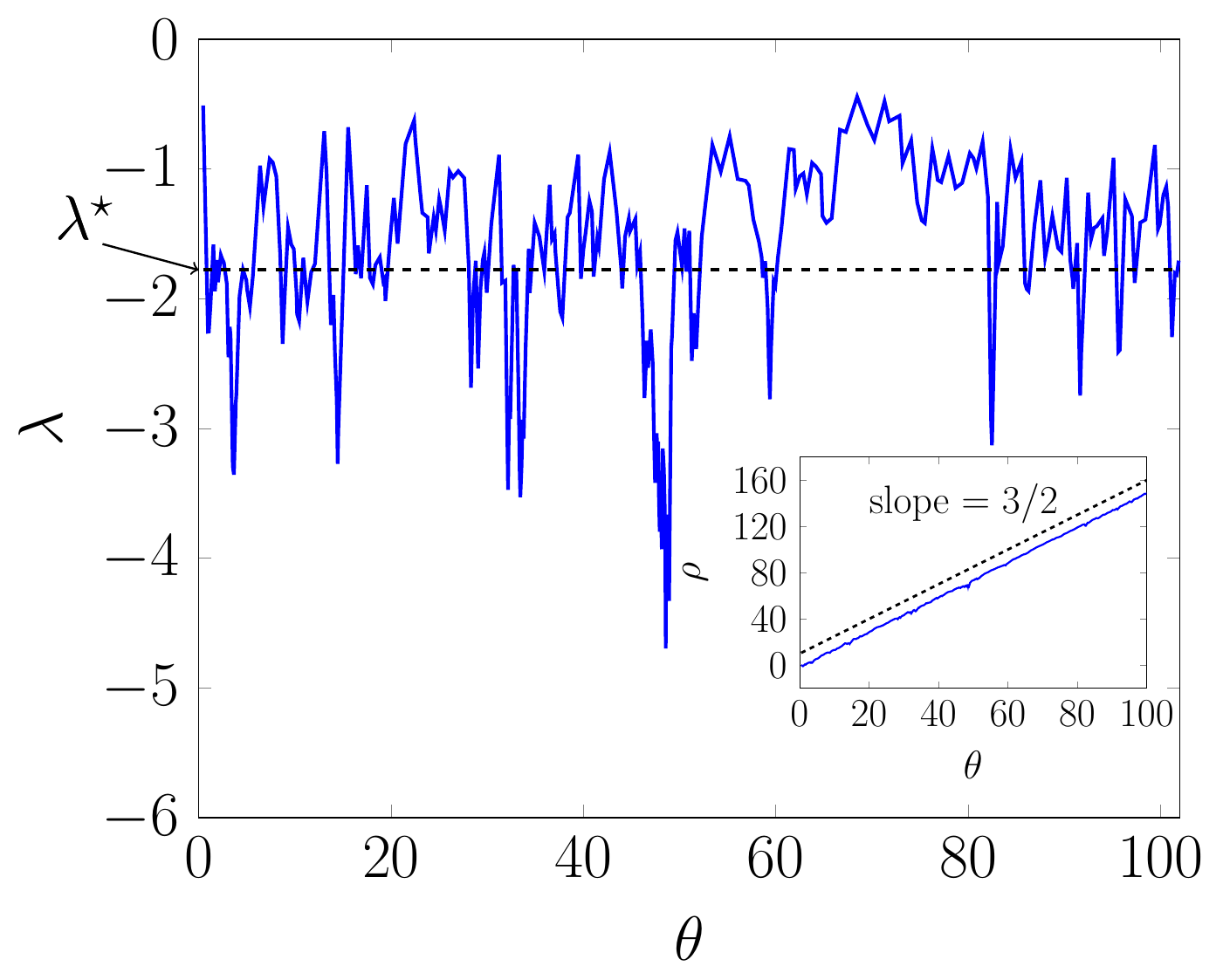}}
\end{minipage}  
\caption{ (Color online) Left: Sketch of a piecewise-ballistic
  scenario. Right: A typical realization of $\lambda$ as a function of the
  physical logarithmic time $\theta$, and the corresponding evolution
  of $\rho = \ln (r/r_0)$ (inset).  Here, $\alpha$ is uniform between
  $-1$ and $1$, and $\log \beta$ is Gaussian with zero mean and unit
  variance (Please see the text for the definitions).  }
\label{fig:ballistic_scheme}
\end{figure}

Given a pair of tracers, we denote $\mathbf r(t)$ and $\delta \mathbf
u(t)$ their separation and relative velocity, respectively.  A
``ballistic modelling'' consists in assuming that during a time-lapse
$\tau$, the velocity $\delta \mathbf u(t)$ remains constant, so that
$\mathbf r(t+\tau) \simeq \mathbf r (t)+ \tau\delta \mathbf u (t)$. As
demonstrated by \cite{bitane-etal:2012} for separations inside the
inertial range, the ballistic motion holds typically for a time $\tau$
of the order of the time needed to damp out $|\delta \mathbf u(t)|^2$
with the average turbulent dissipation rate $\varepsilon$, namely
$\tau \simeq |\delta \mathbf u (t)|^2/(2\varepsilon)$.  Applying
recursively this heuristic argument suggests that the separation
between the two tracers undergoes a sequence of non-correlated
ballistic increases or decreases at times $t_0=0$, $t_1$, $t_2$ \dots
$t_k$ (see Figure \ref{fig:ballistic_scheme} Left). Defining the
origin of time such that $\mathbf r(0)= \mathbf r_0$, the separation
$\mathbf r_k = \mathbf r(t_{k})$, and corresponding time $t_k$ will
then evolve jointly as
\begin{equation}
 \mathbf r_{k+1}=  \mathbf r_k +   \dfrac{|\delta \mathbf
   u_k|^2}{2\varepsilon} \delta \mathbf u_k \hspace{0.5cm} \text{and}
 \hspace{0.5cm}  t_{k+1}=t_k +  \dfrac{|\delta \mathbf
   u_k|^2}{2\varepsilon}.
\label{eq:turningpoints_rt}
\end{equation}
In the language of CTRW, the values $(\mathbf r_k,t_k)$ are called the
\emph{turning points}\/ of the process.  Between two successive
turning points, the separation is by construction ballistic, so that
its value at any time can be obtained by linear interpolation.  This
choice corresponds to a \emph{leaping}\/ CTRW, as already used by
\cite{shlesinger-etal:1987} in the context of a Levy walk description
of turbulent pair dispersion.  To entirely describe the statistics of
the separations, one now only needs to prescribe statistics
(\emph{i}\/) for the moduli of the relative velocities $| \delta
\mathbf u_k|$, and (\emph{ii}\/) for the longitudinal components of
the velocities $\delta \mathbf u_k^\parallel = \delta \mathbf u_k
\cdot \mathbf r_k / | \mathbf r_k |$. In principle the distribution of
turbulent velocity differences non-trivially depends on the scale $|
\mathbf r_k |$. This can be due to finite-size effects, so that one
has to distinguish between dissipative, inertial, and integral
scales. We focus here on separations that always remain within the
inertial range and exclude any intermittent corrections to K41
similarity hypothesis. While crude, such an assumption will prove to
be sufficient to reproduce the main mechanisms of relative
dispersion. In this framework, the relevant inputs are the statistics
of
\begin{equation}
  \alpha_k=\dfrac{\delta \mathbf u_k^\parallel}{|\delta \mathbf u_k|}
  \hspace{0.5cm} \text{and} \hspace{0.5cm} \beta_k
  =\dfrac{|\delta \mathbf u_k|^3}{2\,\varepsilon\,|\mathbf r_k|}.
\end{equation}
The variables $\alpha_k\in[-1,1]$ should have an asymmetric
distribution in order to reproduce the skewness of longitudinal
velocity differences in turbulence. The variables $\beta_k$ account
for the fluctuations in the rate of energy transfer and, under K41
assumptions, should be independent of the $\mathbf r_k$'s. The time
lapses between two consecutive turning points may be thought of as
correlation times: It is then natural to prescribe that both
$\alpha_k$ and $\beta_k$ be independently distributed. We later refer
to the distributions of the noises $\alpha_k$ and $\beta_k$ as
$\alpha$ and $\beta$ ---\,without a subscript\,--- and denote with
$\langle \cdot\rangle$ the average over their realizations.

Under these assumptions the dynamics of the distance $r_k =
|\mathbf{r}_k|$ reduces to
\begin{equation}
  r_{k+1}=  r_k  \left(1+2\,\alpha_k\,\beta_k+\beta_k^2\right)^{1/2}
  \hspace{0.5cm} \text{and}
  \hspace{0.5cm}  t_{k+1}=t_k + (2\,\varepsilon)^{-1/3}\left(\beta_k\, r_k\right)^{2/3}.
\label{eq:turningpoints_ab}
\end{equation}
Note that this process is not purely multiplicative.  As discussed in
the previous section, if the time increments $\tau_k = t_{k+1}-t_k$
were constant, then the distributions of the logarithms of the
distance at a given observation time would evolve towards a Gaussian
distribution given by the Central Limit Theorem.  This is however not
the case here, as the time associated to a given pair of tracers is
itself a random variable, which is neither additive nor
multiplicative.  In the present paper, we do not need to prescribe
further the distributions of $\alpha$ and $\beta$. The only constraint
concerns the quantity $\langle \ln \left( 1 + 2\alpha \beta + \beta^2
\right) \rangle $. It is required to be positive in order to ensure
that the times $t_k$ go to infinity as the number of turning points
diverges. This prevents the sequence $t_k$ from converging and the two
tracers from touching each other in a finite time.

\subsection{Statistics of the separations from the piecewise ballistic perspective}

We shall not here attempt to work out in full mathematical details the
statistics of the separations which the model predicts. Rather, we
focus on a general and qualitative description of those, based on
simple physical arguments.

\subsubsection{Self-Similarity}
The piecewise-ballistic scenario as modeled by the system
\eqref{eq:turningpoints_ab} yields a whole family of self-similar
distributions for the separations. To understand the origin of this
self-similarity, it is convenient to introduce the non-dimensional
``logarithmic'' variables
\begin{equation}
  \theta_k = \ln ({t_k}/{\tau_0}),\quad \rho_k = \ln ({r_k}/{r_0}),
   \tieq{and} \lambda_k = \ln ({r_k}/{r_0}) - ({3}/{2}) \ln ({t_k}/{\tau_0}),
\label{eq:thetalambda_def}
\end{equation}
where $\tau_0 = (2\,\varepsilon)^{-1/3}\,r_0^{2/3}$ is a
characteristic time lapse associated to the initial separation
$r_0$. The largest values of $\lambda$ indicate pairs that separate
faster than the typical explosive separation, while large negative
values are obtained when they get close to each other.  We next define
$p_\theta(\rho)$ the probability density of $\rho$ conditioned on
$\theta$, regardless of the number of turning points needed to reach
this time, and write $\langle\cdot \rangle_\theta$ the corresponding
average.

Self-similarity is achieved if the statistics of the variable
$\lambda_k$ become steady at long times, \emph{e.g.}\/ when $\theta
\gg 1$. A rigorous proof that $\lambda_k$ evolves towards a stationary
distribution $\pstat$ goes beyond the scope of the present
paper. Still, let us give qualitative arguments, starting from the
evolution of $\theta_k$ and $\lambda_k$.  The combination of
\eqref{eq:turningpoints_rt} and \eqref{eq:thetalambda_def} yields
\begin{equation}
  \lambda_{k+1}=\lambda_k +\dfrac{3}{2}\ln \dfrac{\left(1+ 2\alpha_k\beta_k +
      \beta_k^2\right)^{\frac{1}{3}}}{1+\beta_k^\frac{2}{3}{\rm e}^{\frac{2}{3}
      \lambda_k}}
  \tieq{and} 
  \theta_{k+1} =  {\theta_k} + \ln \left( 1 +  \beta_k
    ^{\frac{2}{3}}{\rm e}^{\frac{2}{3} \lambda_k} \right).
\label{eq:turningpoints_lambdatheta}
\end{equation}
The equation for $\lambda_k$ is closed as it does not involve the time
variable $\theta_k$. A typical realization of the $\lambda_k$'s is
shown in Figure~\ref{fig:ballistic_scheme} (Right). They fluctuate
around the specific value $\lambda^\star$, univocally defined by
\begin{equation}
  \left\langle\ln \left(1+ 2\alpha\beta + \beta^2\right)\right\rangle = 3
    \left\langle \ln \left( 1+\beta^{2/3} e^{2
          \lambda^\star/3}\right) \right\rangle.
\end{equation}
It is easily shown that $\lambda^\star$ is always negative. The
evolution of $\lambda_k$ can then be decomposed as the sum of a noise
$\mathcal W$ with zero mean and a restoring stochastic force $\mathcal
F$, namely
\begin{equation}
  \lambda_{k+1} = \lambda_k + \mathcal  W(\alpha_k,\beta_k) + \mathcal  F(\lambda_k,\beta_k), 
  \text{~with~} \left\{\begin{array}{l} \mathcal W (\alpha,\beta)
      = \frac{3}{2}\ln\frac{\left|1+ 2\alpha \beta +
          \beta^2\right|^{1/3}}{ 1+\beta^{2/3} {\rm
          e}^{2\lambda^\star/3}}, \\
      \mathcal  F(\lambda,\beta) = \frac{3}{2}\ln
      \frac{1+\beta^{2/3}{\rm e}^{2\lambda^\star/3}}{1+\beta^{2/3}{\rm
          e}^{2\lambda/3}}. \end{array}\right.
\label{eq:turningpoints_OU}
\end{equation}
The force $\mathcal F(\lambda,\beta)$ is a decreasing function of
$\lambda$ and changes sign at $\lambda=\lambda^\star$. We therefore
expect $\lambda^\star$ to be a recurrent point. This is the source of
the stationarity of the process $\lambda_k$, which we have observed
numerically. The stationarity of $\lambda_k$ has several consequences.
First, the average logarithmic separation grows asymptotically when
$\theta \gg 1$ as $\langle \rho \rangle_\theta = {3}\theta/2 + \langle
\eta \rangle_{\infty} + {\rm o}(1)$ (see inset of Figure
\ref{fig:ballistic_scheme} Right), while its variance becomes
constant, as observed earlier in our data. Second, the stationarity of
$\lambda_k$ implies an explosive behaviour of separations. The initial
separation $r_0$ does not appear in the dynamics
(\ref{eq:turningpoints_lambdatheta}). Also, one can easily see that
the specific choice of $\tau_0$ ensures that $r_0$ can be simplified
in the definition of $\lambda_k$. The dependence upon the initial
separation is thus entailed in the definition of $\theta_k$. However,
this dependence disappears when $t\to\infty$, so that the stationary
distribution of $\lambda_k$ is independent of $r_0$.  Finally, for
large-enough times the probability density of the logarithmic
separations is simply translated by the dynamics around its average
value: $p_\theta(\rho) = \pstat(\rho-3\theta/2)$. This implies that,
at large times, the statistics of the distance $r$ approach the
self-similar form $p(r,t) \simeq r^{-1}
\pstat[\ln(r/(\varepsilon^{1/2}\,t^{3/2}))]$, which is independent of
$r_0$.

\subsubsection{Tails}
The piecewise ballistic mechanism does not yield a single but a whole
family of self-similar distributions $\pstat$, depending on the
choices of $\alpha$ and $\beta$. We can however try to characterize
the tails of $\pstat$.

The right-end tail ($\lambda > \lambda^\star$) corresponds to tracers
that separate faster than the average.  At large times, the piecewise
model \eqref{eq:turningpoints_lambdatheta} predicts that
$\lambda(\theta)$ is negative, or in other words, that the logarithmic
distance $\rho$ is strictly smaller than three halves of the
logarithmic time. It is indeed easily checked that if $\lambda$ is
negative at a given turning point, then it remains negative
afterwards. As $\lambda$ winds up fluctuating around
$\lambda^\star<0$, it is almost surely negative at large times.
A noticeable consequence is that the limiting self-similar
distributions have a right-end cutoff.  Such a behaviour contrasts
that obtained from eddy-diffusivity models (including Richardson's),
whose tails fall as a double exponential at large values.

The other tail ($\lambda \to -\infty$) captures the statistics of
pairs that are \emph{not} separated. Numerical simulations of the
model reveal that its form does strongly depend on the distribution of
$\alpha$ and $\beta$. To understand further this behaviour, let us
define $\Gamma= \langle \mathcal F (-\infty,\beta) \rangle$ and
$\kappa = \langle \mathcal W(\alpha,\beta)^2\rangle^{1/2}$, and
describe two asymptotic regimes $\kappa \ll \Gamma$ and $\kappa \gg
\Gamma$.\\
(\emph{i}\/) $\kappa \ll \Gamma$ describes a situation where tracers
almost never come close to each other. When they do so, they are
immediately pulled back to their initial separation by the restoring
force $\mathcal F$. In this case, one-step excursions dominate the
left-end tail of the statistics, which is therefore entirely
determined by the distribution of $\mathcal
W$.\\
(\emph{ii}\/) The case $\Gamma \ll \kappa$ is opposite. Here, two
particles need to undergo a large number of ballistic steps to be
pulled back towards $\lambda^\star$. The time of each step is in
average smaller than typical correlation time $ \Delta\theta^\star =
\langle \log (1 + \beta^{2/3}{\rm e}^{2\lambda^\star/3})\rangle$ at
$\lambda = \lambda ^\star$. In this limit, the noise is dominant and
the discrete dynamics can be approximated by a Brownian motion with a
positive drift. One can check that this yields a stationary
distribution whose left-end tail is $\propto {\rm
  e}^{(2/3+\gamma)\,\lambda}$ where $\gamma>0$ depends on the noise
kernels. With an accurate choice of the noises, the left exponential
slope is therefore likely to be shallower than Richardson's left
exponential slope of $3$, as suggested by the numerical data.

\section{Scaling of the distribution of distances}
\label{sec:scaling}
At this point, one may wonder whether the piecewise-ballistic
phenomenology achieves a better description of the full statistics
than Richardson's distribution. As the precise shape of the statistics
obtained from the piecewise-ballistic scenario are noise-dependent, a
detailed answer would require to plug into our stochastic model some
``realistic noises'' for the distributions of $\alpha$ and $\beta$.
This goes beyond our intention, since we consider here the set of
equations (\ref{eq:turningpoints_rt}) as a ``scenario'' rather than a
genuine ``modelling'' of the separations.
 
\begin{figure}
  \centerline{\includegraphics[width=0.7\textwidth]{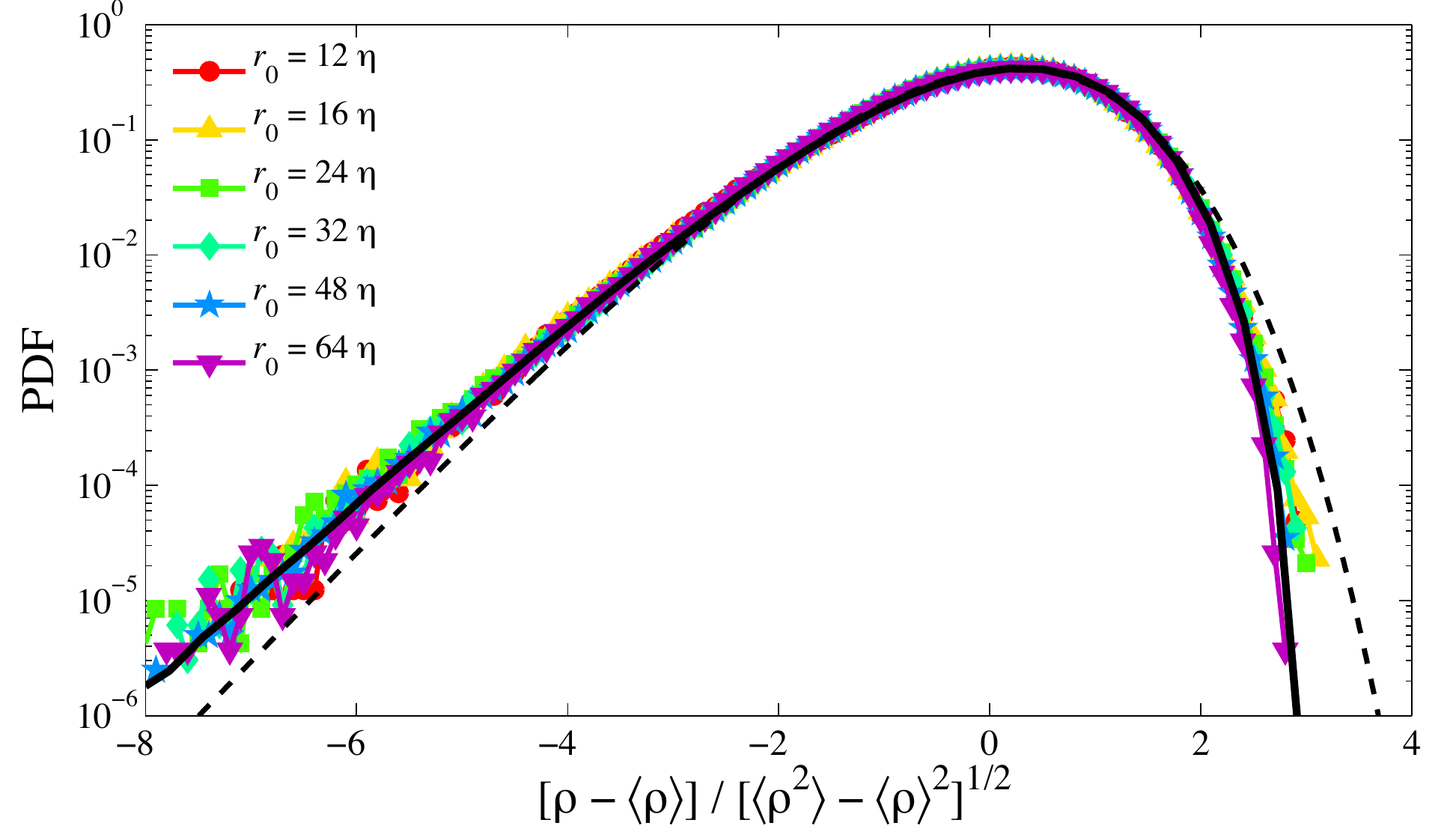} }
  \caption{(Color online) Probability density function of the
    logarithm of the inter-particle distance $\rho = \ln (r(t)/r_0)$
    for various $r_0$ and, in each case, at time $t\approx
    9\,\tau_{r_0}$ with $\tau_{r_0} =
    \varepsilon^{-1/3}r_0^{2/3}$. The dashed line is Richardson's
    distribution. The black solid line is the steady distribution
    obtained numerically from the ballistic model for $\ln \beta \sim
    \mathcal N(0,1)$ and $p(\alpha) = (5/6)
    \left((\alpha+1)/2\right)^{-1/3}$.
  }
\label{fig:collapse_distrib_lnR}
\end{figure}

To our taste, the main virtue of the model is that it provides a
non-Markovian physical interpretation to the explosive nature of the
inter-particle separations, \emph{viz.}\/, the independence upon the
initial separation.  We believe that dynamical memory effects are
essential for the understanding of extreme events in relative
dispersion. One may therefore wonder whether such effects are
universal and whether their signature is intrinsically linked to the
explosive separation. A related question concerns the origin of the
observed deviations to Richardson's self-similar distribution. Are
they due to finite-size effects, to the intermittency of velocity
statistics, or rather, as we think, to the limits of the
eddy-diffusivity approach?  Our impression is that this question has
been somehow overlooked in the previous literature.  Experimental and
numerical datasets are often confronted to Richardson's distribution
using a linlog representation of the distance neighbour function
$p(r,t)/(4\pi\,r^2)$ in such a way that Richardson's distribution
appears as a straight line~\citep[see,
\emph{e.g.}\/,][]{ouellette-etal:2006}. Such a representation puts a
visual emphasis on the collapse of the bulk of the distribution, but
is not optimal for a thorough study of the tails.  In Figure
\ref{fig:collapse_distrib_lnR}, we use linlog coordinates to represent
the distributions of the rescaled logarithmic distances $\tilde \rho
=[{\rho -\langle \rho\rangle }]/[{(\langle\rho^2 \rangle -\langle
  \rho\rangle)^{1/2}}]$ observed in the numerical simulation for six
different initial separations $r_0$ inside the inertial range and for
each at the fixed time $t \approx 9\, \tau_{r_0}$.
With such a choice, the distribution of the logarithmic distances
seems to converge towards a single distribution, regardless of $r_0$.
The collapse of the \emph{full} distribution makes the explosive
nature of pair separation explicit. It is also once again clear that
the limiting distribution is \emph{not} Richardson's (dashed line).
Note that a casual choice for the statistics of $\alpha$ and $\beta$
makes the piecewise ballistic steady distribution (solid line) fit the
data better.  The model predicts a sharp cutoff at large
values. However, the current statistical accuracy does not enable us
to discriminate between such a behaviour and the double exponential
obtained in the framework of eddy-diffusivity approaches.

To conclude, let us stress again that the piecewise-ballistic
phenomenology provides a new and intuitive way of thinking about the
problem of pair dispersion and reproduces some salient statistical
features of tracer separation.  While it might also be used to
investigate possible effects of the fluid flow intermittency, we
limited here our study to the K41 framework. The proposed toy model
displays a number of general trends that include (\textit{i}\/) the
explosive nature of the statistics, or in other words the property
that the steady distributions do not depend on the initial separation;
(\textit{ii}\/) their self-similarity, which makes the statistics of
the logarithm of the separation collapse towards a single
distribution; (\textit{iii}\/) the presence of a right-end cutoff in
the associated probability density; (\textit{iv}\/) the growth of the
average of the logarithmic separation as three halves of the
logarithmic time, compatible with the $t^3$ law, resulting from the
multiplicative nature of the separation process.

\medskip

This research has received funding from the European Research Council
under the European Community's Seventh Framework Program
(FP7/2007-2013 Grant Agreement No.~240579) and from the French Agence
Nationale de la Recherche (Programme Blanc ANR-12-BS09-011-04).

\bibliographystyle{jfm}

\end{document}